\documentclass[conference,10pt]{IEEEtran}
\ifCLASSINFOpdf
\else
\fi

\usepackage{stfloats}
\usepackage{makeidx}  

\usepackage{graphicx}
\usepackage{verbatim}
\usepackage{amsmath}
\usepackage{algorithm}
\usepackage{algpseudocode}
\usepackage{algorithmicx}
\usepackage{tabularx}
\usepackage{multirow}
\usepackage{hhline}
\usepackage{color}

\usepackage[nolist]{acronym}
\begin{acronym}
	\acro{AAA}{Authentication, Authorization and Accounting}
	\acro{AP}{Access Point}
	\acro{API}{Application Programming Interface}
	\acro{BP}{Baseline Pseudonym}
	\acro{BS}{Base Station} 
    \acro{BSM}{Basic Safety Message}
    \acro{BYOD}{Bring Your Own Device}
	\acro{C2C-CC}{Car2Car Communication Consortium}
	\acro{CA}{Certification Authority}
	\acrodefplural{CA}{Certification Authorities}
	\acro{CN}{Common Name}
	\acro{CAM}{Cooperative Awareness Message}
	\acro{CIA}{Confidentiality, Integrity and Availability}
	\acro{CRL}{Certificate Revocation List}
	\acro{CDN}{Content Delivery Network}
	\acro{CSR}{Certificate Signing Requests}
	\acro{DAA}{Direct Anonymous Attestation}
	\acro{DDoS}{Distributed Denial of Service}
	\acro{DDH}{Decisional Diffie-Helman}
	\acro{DENM}{Decentralized Environmental Notification Message}
	\acro{DHT}{Distributed Hash Table}
	\acro{DoS}{Denial of Service}
	\acro{DPA}{Data Protection Agency}
	\acro{ECU}{Electronic Control Unit}
	\acro{ETSI}{European Telecommunications Standards Institute}
	\acro{ECDSA}{Elliptic Curve Digital Signature Algorithm}
	\acro{ECC}{Elliptic Curve Cryptography}
	\acro{EVITA}{E-safety Vehicle Intrusion protected Applications}
	\acro{FCFS}{First-Come First-Served}
	\acro{FOT}{Field Operational Test}
	\acro{FPGA}{Field-Programmable Gate Array}
	\acro{GN}{GeoNetworking}
	\acro{GS}{Group Signatures}
	\acro{GM}{Group Manager}
	\acro{GBA}{Generic Bootstrapping Architecture}
	\acro{GPA}{Global Passive Adversary}
	\acro{GUI}{Graphic User Interface}
	\acro{HP}{Hybrid Pseudonym}
	\acro{HSM}{Hardware Security Module}
	\acro{HTTP}{Hypertext Transfer Protocol}
	\acro{IEEE}{Institute of Electrical and Electronics Engineers}
	\acro{IoT}{Internet of Things}
	\acro{ITS}{Intelligent Transport Systems}
	\acro{IT}{Information Technologies}
	\acro{IMSI}{International Mobile Subscriber Identity}
	\acro{IMEI}{International Mobile Station Equipment Identity}
	\acro{IdP}{Identity Provider}
	\acro{ISP}{Internet Service Provider}
	\acro{LEA}{Law Enforcement Agency}
	\acro{LTC}{Long-Term Certificate}
	\acro{LTCA}{Long-Term \acl{CA}}
	\acrodefplural{LTCA}{Long-Term \aclp{CA}}
	\acro{LDAP}{Lightweight Directory Access Protocol}
	\acro{LTE}{Long Term Evolution}
	\acro{LBS}{Location-based Service}
	\acro{MAC}{Message Authentication Code}
	\acro{MCA}{Message \ac{CA}}
	\acro{MEA}{Misbehavior Evaluation Authority}
	\acro{OBU}{On-Board Unit}
	\acro{OCSP}{Online Certificate Status Protocol}
	\acro{OSN}{Online Social Network}
	\acro{PC}{Pseudonymous Certificate}
	\acro{PCA}{Pseudonymous \acl{CA}}
	\acrodefplural{PCA}{Pseudonymous \aclp{CA}}
	\acro{PDP}{Policy Decision Point}
	\acro{PEP}{Policy Enforcement Point}
	\acro{PIR}{Private Information Retrieval}
	
	\acro{PKI}{Public-Key Infrastructure}
	\acro{POI}{Point of Interest}
	\acro{PRECIOSA}{Privacy Enabled Capability in Co-operative Systems and Safety Applications}
	\acro{PRESERVE}{Preparing Secure Vehicle-to-X Communication Systems}
	\acro{P2P}{peer-to-peer}
	\acro{RA}{Resolution Authority}
	\acro{REST}{Representational State Transfer}
	\acro{RBAC}{Role Based Access Control}
	\acro{RCA}{Root \acl{CA}}
	\acro{RSU}{Roadside Unit}
	\acro{SAML}{Security Assertion Markup Language}
	\acro{SCORE@F}{Système COopératif Routier Expérimental Français}
	\acro{SeVeCom}{Secure Vehicle Communication}
	\acro{SIS}{Security Infrastructure Server}
	\acro{SP}{Service Provider}
	\acro{SSO}{Single-Sign-On}
	\acro{SoA}{Service-oriented-Approach}
	\acro{SOAP}{Simple Object Access Protocol}
	\acro{SAS}{Sample Aggregation Service}
	\acro{TS}{Task Service}
	\acro{TLS}{Transport Layer Security}
	\acro{TPM}{Trusted Platform Module}
	\acro{TTP}{Trusted Third Party}
	\acro{TVR}{Ticket Validation Repository}
	\acro{URI}{Uniform Resource Identifier}
	\acro{VANET}{Vehicular Ad-hoc Network}
	\acro{V2I}{Vehicle-to-Infrastructure}
	\acro{V2V}{Vehicle-to-Vehicle}
	\acro{V2X}{\ac{V2V} and/or \ac{V2I}}
	\acro{VC}{Vehicular Communication}
	\acro{VM}{Virtual Machine}
	\acro{VSN}{Vehicular Social Network}
	\acro{VSS}{\ac{VC} Security Subsystem}
	\acro{WAVE}{Wireless Access in Vehicular Environments}
	\acro{WSDL}{Web Services Discovery Language}
	\acro{W3C}{World Wide Web Consortium}
	\acro{VANET}{Vehicular Ad-hoc Network}
	\acro{VPKI}{Vehicular Public-Key Infrastructure}
	\acro{WS}{Web Service}
	\acro{WoT}{Web of Trust}
	\acro{WSACA}{\ac{WAVE} Service Advertisement \ac{CA}}
	\acro{XML}{Extensible Markup Language}
	\acro{XACML}{eXtensible Access Control Markup Language}
	\acro{3G}{3rd Generation}
\end{acronym}

\usepackage{algpseudocode}
\usepackage{varwidth}
\usepackage{subcaption}

\usepackage{algorithm}
\PassOptionsToPackage{noend}{algpseudocode}
\usepackage{algpseudocode}

\errorcontextlines\maxdimen

\makeatletter
\newcommand*{\algrule}[1][\algorithmicindent]{\makebox[#1][l]{\hspace*{.5em}\vrule height .75\baselineskip depth .25\baselineskip}}%

\newcount\ALG@printindent@tempcnta
\def\ALG@printindent{%
	\ifnum \theALG@nested>0
	\ifx\ALG@text\ALG@x@notext
	\addvspace{-3pt}
	\else
	\unskip
	\ALG@printindent@tempcnta=1
	\loop
	\algrule[\csname ALG@ind@\the\ALG@printindent@tempcnta\endcsname]%
	\advance \ALG@printindent@tempcnta 1
	\ifnum \ALG@printindent@tempcnta<\numexpr\theALG@nested+1\relax
	\repeat
	\fi
	\fi
}%
\usepackage{etoolbox}
\patchcmd{\ALG@doentity}{\noindent\hskip\ALG@tlm}{\ALG@printindent}{}{\errmessage{failed to patch}}
\makeatother

\begin{document}
	%
	\title{Scaling VANET Security Through Cooperative Message Verification}

	\author{\IEEEauthorblockN{Hongyu Jin and Panos Papadimitratos}
	\IEEEauthorblockA{Networked Systems Security Group, KTH Royal Institute of Technology, Sweden \\
		\emph{\{hongyuj, papadim\}}@kth.se\\
		www.ee.kth.se/nss}}


	\maketitle
	
	\begin{abstract}
		VANET security introduces significant processing overhead for resource-constrained \acp{OBU}. Here, we propose a novel scheme that allows secure \ac{VC} systems to scale well beyond network densities for which existing optimization approaches could be workable, without compromising security (and privacy).
	\end{abstract}
	

	%
	\IEEEpeerreviewmaketitle

\section{Introduction}

\acf{VC} systems, notably \ac{V2V} and \ac{V2I} communication, entail high-rate transmissions; typically, for safety applications, \acfp{OBU} transmit at a rate of 10 messages (safety beacons) per second. Granted, there are methods that adapt the beaconing rate, but the challenge is clear: often, especially as \ac{VC} systems get progressively widely deployed, each vehicle will have to process safety beacons, along with other traffic, from several tens of other vehicles within its \ac{OBU} range. For example, 200 (300) messages/second for 20 (30) neighboring vehicles.

The provision of security and privacy protection aggravates the situation, adding communication overhead (digital signatures and certificates attached, thus longer messages) as well as computational overhead (digital signature verifications mostly, and signature calculations). This problem has been investigated in the literature, with a number of improvements (e.g.,~\cite{calandriello2011performance,feiri2014formal}) compatible with the standardized pseudonymous authentication approach. For example, the certificates of the authenticating sender can be omitted periodically or based on the sender's context~\cite{feiri2014formal}; at the receiver side, they need to be validated once~\cite{calandriello2011performance}. 

These approaches provide significant improvements and show how one can dimension processing power~\cite{calandriello2011performance}, but they are somewhat conservative: they assume each node validates all received messages it receives and deems relevant. Indeed, this is the straightforward approach. An alternative, adaptive, reactive approach has been considered in \cite{ristanovic2011adaptive}, but only for multi-hop messages. 

It is important to realize that for safety beacon transmitted, there are $N$ validations that take place; where $N$ is the number of receiving neighbors. This is significant redundant effort, especially when most of the transmissions, termed \acp{CAM}, are important yet not of high priority. Intuitively, if most nodes (vehicles) are benign, each validation of a message they perform could serve their peers in the vicinity. 

This is exactly the idea we promote in this short paper: Each node can notify its neighbors about its successful verification of some recently received beacons; each neighbor that validates such an augmented message can leverage this additional information and avoid verifying itself the corresponding beacons. Ideally, this could reduce significantly the overhead, as one costly cryptographic verification provides information for multiple messages. But this is a double-edged sword: if not done carefully, it leaves space for abuse by intelligent internal adversaries. 

Our results show that the network density, which raises the scalability problem, can also provide a remedy. Our scheme trades a tiny window of vulnerability, allowing a miniscule fraction of beacons that could be mistaken as valid before an adversarial node be evicted.\footnote{We emphasize that if a message is critical, the receiver can always validate it in a traditional way, in addition to our optimistic approach.} At the same time, we get extensive improvement in terms of message validation delay, allowing, in fact, the network to scale to neighborhoods that are 50\% to 100\% larger than the ones for which optimized yet more conservative approaches would be saturated and unworkable. 

In the rest of the paper, we present in detail our cooperative validation scheme (Sec.~\ref{sec:our-scheme}), we provide a security analysis (Sec.~\ref{sec:analysis}), and present a body of simulation results (Sec.~\ref{sec:evaluation}) before some concluding remarks (Sec.~\ref{sec:conclusions}).  

\section{Our Scheme}\label{sec:our-scheme}

\textbf{Overview:} Our scheme extends the traditional \ac{V2V} message verification, leveraging neighboring peers to reduce validation delays without compromising the achieved security (and privacy). The basic idea is to augment each (safety) message with brief identifiers of previously validated messages. These identifiers indicate the corresponding messages have been verified by the sender. This is exactly where nodes can benefit from each other: accepting a message can help verifying the messages (received and queued) the identifiers in this message point to. In addition, to counter misbehavior, each node probabilistically selects a subset of the received identifiers and verifies by itself the signatures of the corresponding messages. Revocation would be triggered if any misbehavior is identified. Table~\ref{table:notation} summarizes notation used in this paper.

\begin{table}[htp]
	\caption{Notation}
	\centering
	\renewcommand{\arraystretch}{1.10}
	\begin{tabular}{l | *{1}{c} r}
		\hline \hline
		$N$ & \emph{Number of vehicles} \\\hline
		$\{msg\}_\sigma$ & \emph{Signed message} \\\hline
		$Pr_{check}$ & \emph{Probability of checking each peer-provided verification result} \\\hline
		$\alpha$ & \emph{Number of verification results in a CAM} \\\hline
		$\gamma$ & \emph{CAM frequency} \\\hline
		$\tau$ & \emph{Average message verification delay} \\\hline
		$H()/H$ & \emph{Hash function/Hash value} \\\hline
		$b$ & \emph{1 bit value, indicating the message is selected for checking} \\\hline
		\hline
	\end{tabular}
	\renewcommand{\arraystretch}{1}
	\label{table:notation}
\end{table}

\textbf{Message Generation and Reception:} The format of a signed \ac{CAM} in our scheme is changed into:
\begin{align}
\{M\}_{\sigma} = \{CAM \ Fields, H_1..H_\alpha\}_{\sigma}.
\end{align}

Except the hashes, $H_1..H_\alpha$, we assume the rest of the fields are as defined in the standard~\cite{cam}. $H_1..H_\alpha$ are the hashes of latest verified \acp{CAM} (based on the timestamps of the \acp{CAM}, not the times of reception or verification). For example, a vehicle, $V$, caches locally the hash values of the latest verified \acp{CAM} (for which $V$ performed signature verifications, not cooperatively verified as described below) and includes them in its own (sent) \acp{CAM}. 

Everytime a node receives a \ac{CAM}, it generates a job based on the \ac{CAM}. Here, we consider the processing of a received \ac{CAM} as a job. The format of the job is defined as follows:
\begin{align}
\{\{M\}_{\sigma}, H(\{M\}_{\sigma}), b=0 \ or \ 1\}.
\end{align}

The field $b$ indicates the \acp{CAM} are selected for probabilistic checking. In case $b$ is set to 1 for a job, the corresponding \ac{CAM} cannot be verified through cooperative verification (peer-provided hashes): the signature of this \ac{CAM} must be verified. For each new job, $b$ is set to 0.

We assume a single thread for cryptographic verification in each \ac{OBU}: a \ac{CAM} received when the thread is busy needs to be queued. To increase efficiency, we \emph{randomly select} the inserted position in the queue for each new job. This way, we reduce the probability that nearby receivers verify the same \ac{CAM} roughly simultaneously. The verification of \acp{CAM} that sent from a node does not need to follow the sending order, as long as they are verified before they expire. 



\textbf{Cooperative Verification:} Queue processing at a node is done according to Algorithm~\ref{cooperative_verification}. When the queue is not empty, the node pops the first job from the queue, and accepts the \ac{CAM} if the signature is valid. The hashes, $H_1..H_\alpha$, are used to verify the \acp{CAM} (for which $b=0$) in the queue. For each cooperatively verified \ac{CAM}, there is a probability $Pr_{check}$, that the \ac{CAM} will be checked by validating the signature. If so, $b$ is set to $1$ and it is inserted after the last job with $b=1$ (i.e., before the first job that $b=0$). Otherwise, the \ac{CAM} is accepted and removed from the queue.

The probabilistic checking of the claimed verifications (hashes) counters abuse, due to (i) the density of the neighborhood and the (extensive, most often) majority of benign nodes present, and (ii) the ability to locally contain/ignore misbehaving nodes and then globally evict them. The latter is easily enabled by our scheme (simple cryptographic validation in lieu of misbehavior detection), yet the exact way to identify locally the wrongdoer and evict it is orthogonal and can be done by schemes proposed in the literature, e.g.,~\cite{moore2008fast}.
\begin{algorithm}[htp]
	\caption{Cooperative verification}
	\label{cooperative_verification}
	\small 
	\begin{algorithmic}[1]
		\While {$Queue$ is not empty}
			\State Pop a job, $\{\{M\}_{\sigma}, H, b\}$, from the head of $Queue$
			\State $M = \{CAM \ Fields, H_1..H_\alpha\}$
			\If {The signature of $\{M\}_{\sigma}$ is valid}
				\State Accept $M$
				\For{Each $H_i$ in $H_1..H_\alpha$ of $M$}
					\If {$H_i$ is found in $Queue$, and $b$ of $M_i$ is $0$}
						\State Chooses $1$ with probability $Pr_{check}$,
						\State  or $0$ with probability $1- Pr_{check}$
						\If {Chooses $1$}
							\State Insert $\{\{M\}_{\sigma}, H(M), b=1\}$ into $Queue$,
							\State right after the last job, for which $b$ is $1$
						\Else
							\State Accept $M_i$,
							\State and remove $\{\{M_i\}_{\sigma}, H_i, b\}$ from $Queue$
						\EndIf
					\EndIf
				\EndFor
		\EndIf
		\EndWhile
	\end{algorithmic}
\end{algorithm}





\section{Security Analysis}\label{sec:analysis}

\textbf{Adversary Model:} We consider internal adversaries seeking to abuse the system, in particular the cooperative verification scheme. Without loss of generality, let an adversary controlling one \ac{OBU} injecting bogus, i.e., \emph{not properly signed} messages. Then, let the adversary use a compromised \ac{OBU} (equipped with the appropriate credentials) that transmits augmented messages falsely claiming previously transmitted bogus messages as verified. In the hope that those bogus messages received by benign nodes would be accepted without verification/checking. Such adversarial behavior is actually relevant to the optimistic cooperative validation approach we advocate here. An attacker could try to generate malicious \acp{CAM} given overheard hashes. However, given the properties of hash functions and the length of hash values (80 $bit$, as considered in Sec.~\ref{sec:evaluation}), it is very straightforward that finding a message based on the hash values is very hard.

\textbf{Analysis:} Next, we discuss exactly how this misbehavior, specific to our scheme, is countered. We emphasize we are not concerned here with the validity of the message content, e.g., the correctness of a location or an alert about emergency braking; those are orthogonal and can be addressed by relevant consistency checking (\cite{festag2010design, leinmuller2006improved}) and data-centric security schemes \cite{raya2008data}. Here, we are concerned with incorrectly signed (with arbitrary content) messages, and the attempt to legitimize them by improper, adversarial use of our scheme. The detection of any such false claim is straightforward for any legitimate receiver, as long as it cryptographically validates the purported as verified message (signature).

The use of pseudonymous authentication, as per the standards under development, guarantees \emph{non-repudiation} and \emph{message integrity and authentication}; as long as the receiving node performs the cryptographic validation itself (message signature and attached pseudonym validation). 




\textbf{Revealing False Claims:} To increase the probability of detecting such misbehaviors, the reasonable amount of peer-provided verification results should be checked. However, cooperation among nodes within a neighborhood could significantly increase this probability.


Consider a single adversary case, transmitting messages at a rate $\gamma_{adv}$. In the worst case (broadcasting aggressively bogus messages, hoping benign nodes receive as many bogus messages as possible), the adversary could broadcast bogus messages at a rate $\frac{\alpha}{\alpha+1} \cdot \gamma_{adv}$ and broadcast valid messages that ``validate'' those bogus messages at a rate $\frac{1}{\alpha+1} \cdot \gamma_{adv}$. More specifically, broadcast $\alpha$ bogus messages and broadcast the $\left( \alpha + 1 \right)$-th, as a valid one that includes hashes of the earlier $\alpha$ bogus messages. We seek to detect the adversary that provided such faulty claims and revoke its credentials. For any message, the probability of all $\alpha$ included hashes not being checked by a receiver is:
\begin{align}
Pr_{skip} = (1-Pr_{check})^\alpha. \label{eq:skip}
\end{align}

Then, assuming $v$ votes (misbehavior reports) are needed to cooperatively reveal a misbehavior, the probability of revealing such a message with $N$ benign nodes in the neighborhood (assuming they have all received the bogus ``validating'' message) can be estimated as:
\begin{align}
Pr_{reveal} = 1 - \sum_{i=0}^{v-1}\binom{N}{i}(Pr_{skip})^{N-i}(1-Pr_{skip})^i,
\end{align}
where $Pr_{skip}$ is calculated with Eq.~(\ref{eq:skip}). $Pr_{reveal}$ increases with neighbor density; high neighbor density (thus, high message reception rate) environments are the ones our scheme fits best. Note that $Pr_{reveal}$ is only the probability of revealing a single malicious message. The probability of revealing misbehavior after $n$ malicious messages are sent out would be even higher: $1- (1 - Pr_{reveal})^n$.

For example, $Pr_{reveal}$ is around $0.80$ in a network with $\alpha=5$, $N=15$, $Pr_{check}=0.1$ and $v=5$. This means the adversary would be revealed with high probability even after its first transmission of a false claim. A more intelligent adversary could include only a reduced number of bogus message hashes in each valid \ac{CAM}, to reduce the probability of getting detected. However, this significantly weakens the adversary, which would still get revealed after several rounds.

Benign receivers may consume only a few bogus messages for the short period, causing minimal harm while ensuring integrity and authentication of the vast majority of messages in an efficient way; considering most of the \acp{CAM} are sent out with low priority in a usual environment. In addition, a node can choose to verify the \acp{CAM} sent out with high priority immediately, ignoring the queue size and our cooperative verification protocol.

\textbf{Privacy:} On top of security, we note that privacy is not weakened by our scheme. The hash values in a \ac{CAM} do not link transmissions of any other node, beyond what one can infer from geographical information included in the \acp{CAM}.


\section{Evaluation}
\label{sec:evaluation}

\begin{figure*}
	\centering
	\begin{subfigure}[b]{.65\columnwidth}
		\includegraphics[width=\columnwidth]{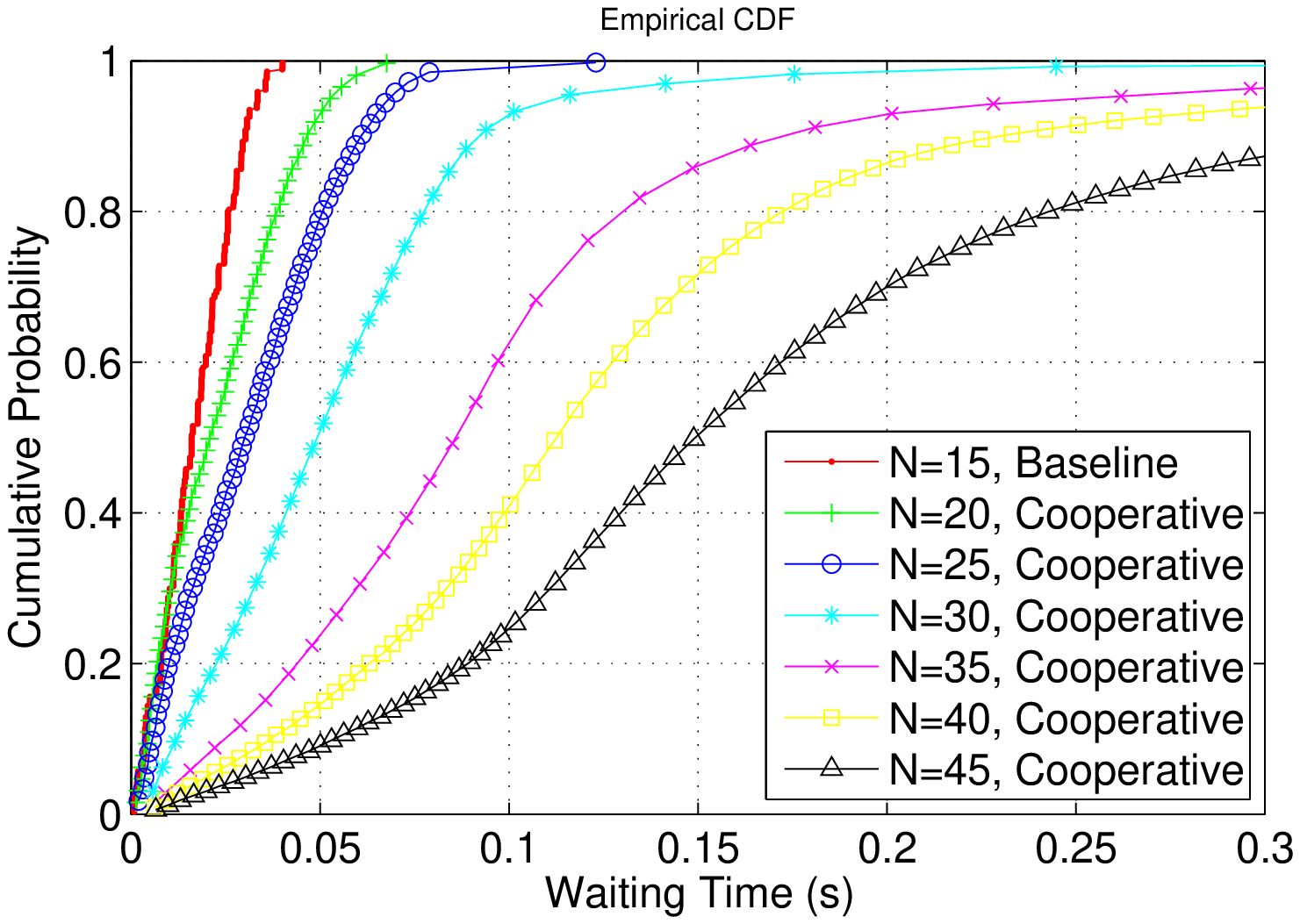}%
		\caption{}%
		\label{subfig1a}%
	\end{subfigure} 
	\begin{subfigure}[b]{.65\columnwidth}
		\includegraphics[width=\columnwidth]{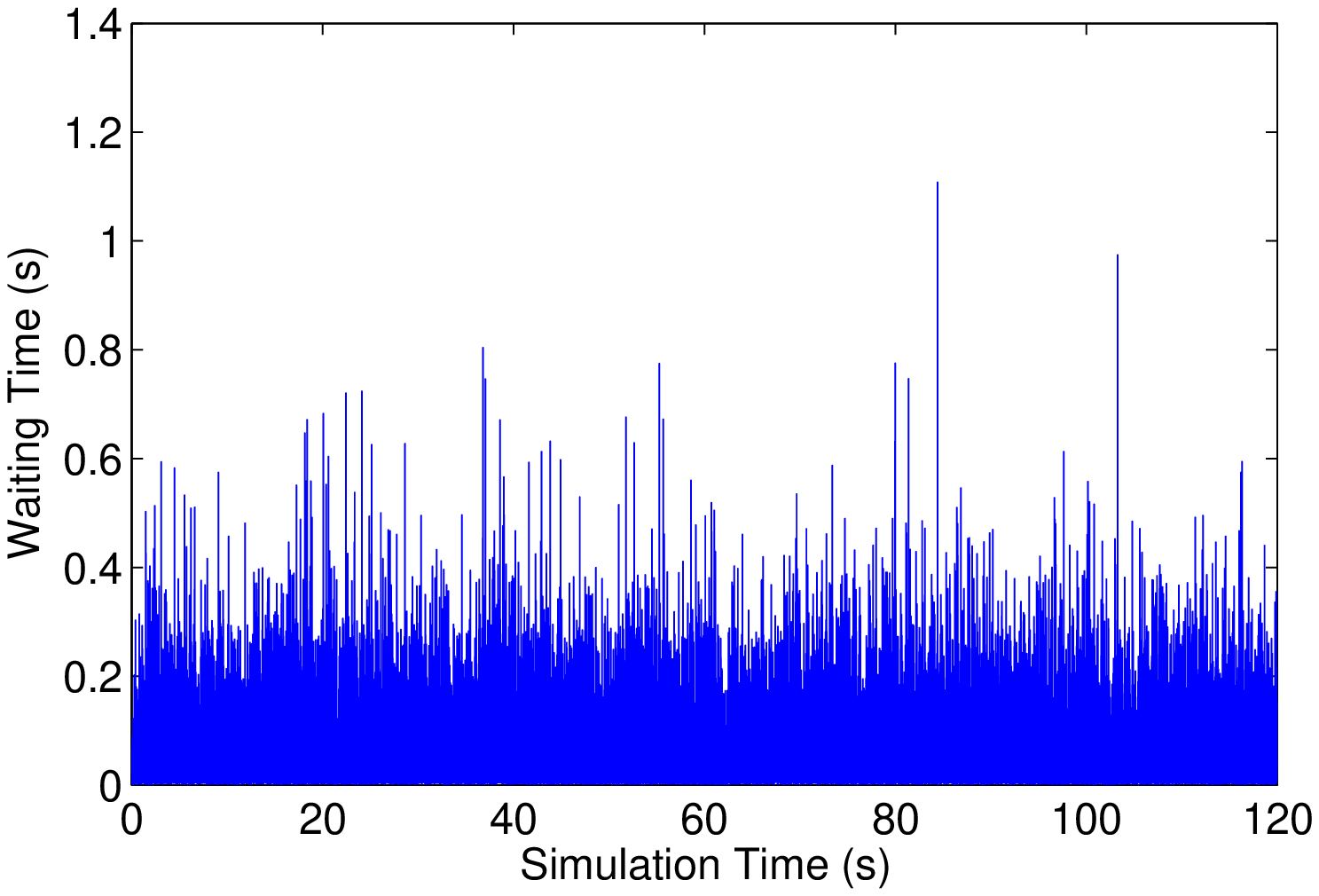}%
		\caption{}%
		\label{subfig1b}%
	\end{subfigure}%
	\begin{subfigure}[b]{.65\columnwidth}
		\includegraphics[width=\columnwidth]{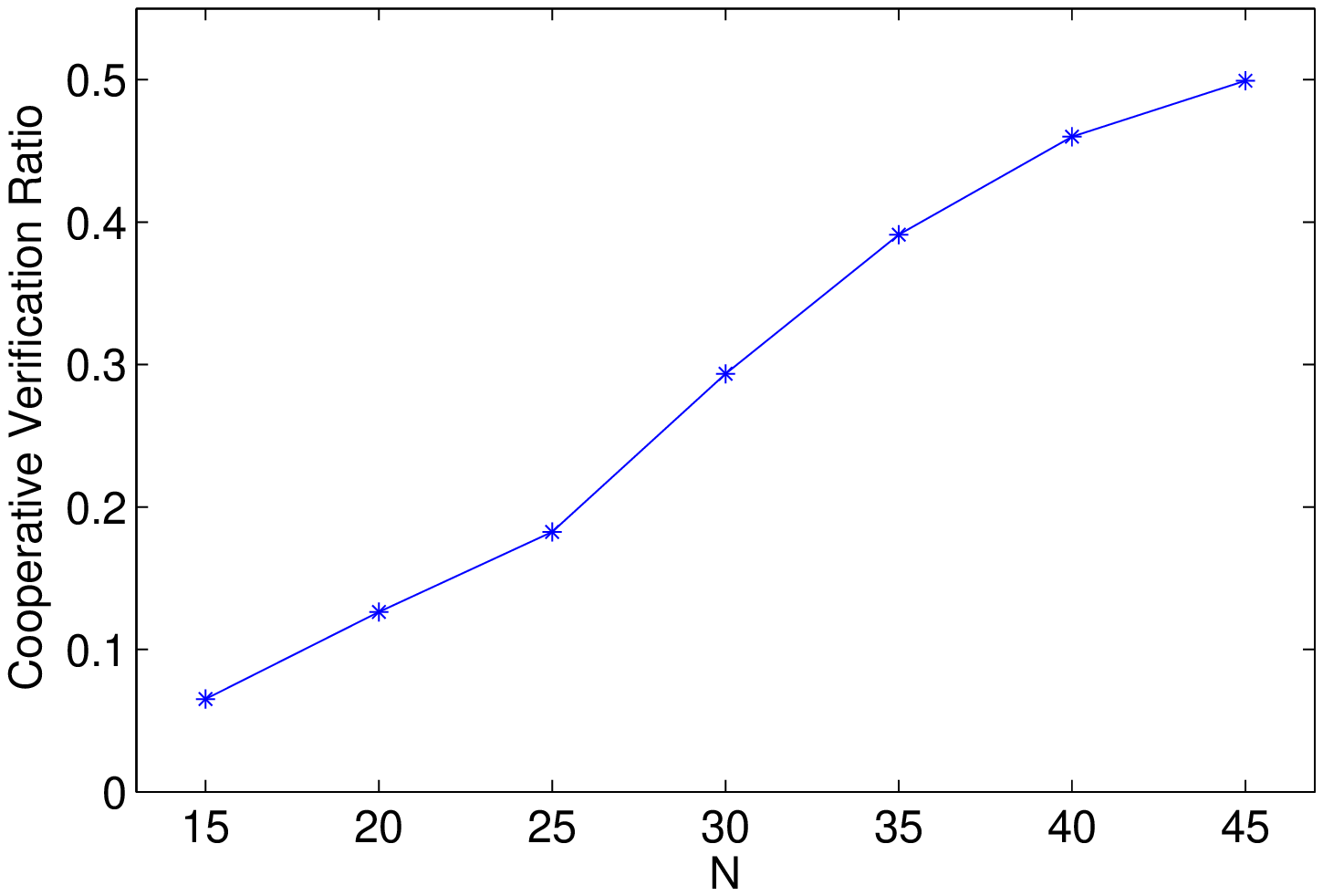}%
		\caption{}%
		\label{subfig1c}%
	\end{subfigure} 
	
	\caption{(a) Waiting time CDF as a function of $N$. (b) Waiting time as simulated time progresses. (Default setting) (c) Cooperative message verification ratio as a function of $N$.}
	\label{fig1abc}
\end{figure*}
\begin{figure*}
	\centering
	\begin{subfigure}[b]{.65\columnwidth}
		\includegraphics[width=\columnwidth]{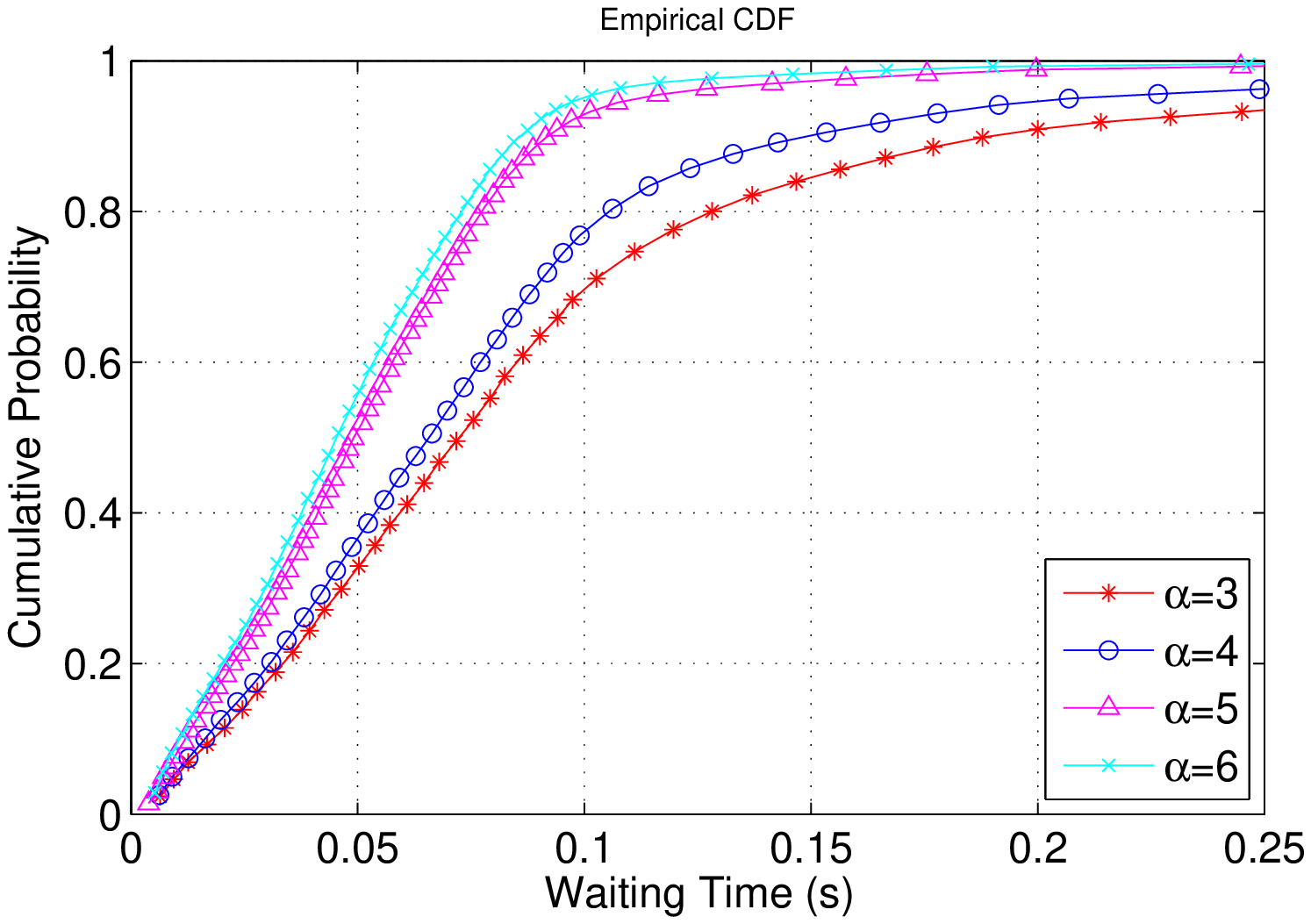}%
		\caption{}%
		\label{subfig2a}%
	\end{subfigure} 
	\begin{subfigure}[b]{.65\columnwidth}
		\includegraphics[width=\columnwidth]{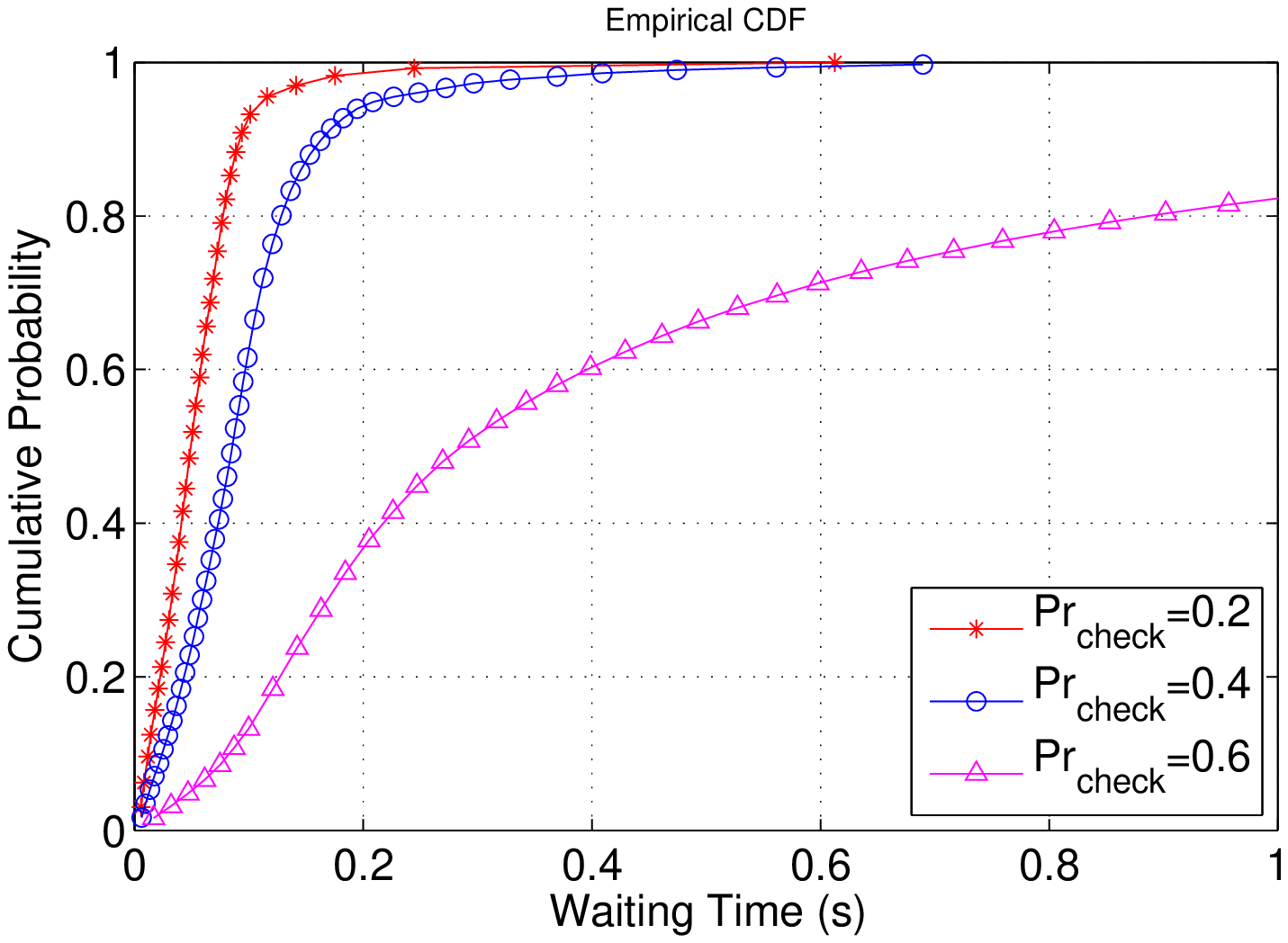}%
		\caption{}%
		\label{subfig2b}%
	\end{subfigure} 
	\begin{subfigure}[b]{.65\columnwidth}
		\includegraphics[width=\columnwidth]{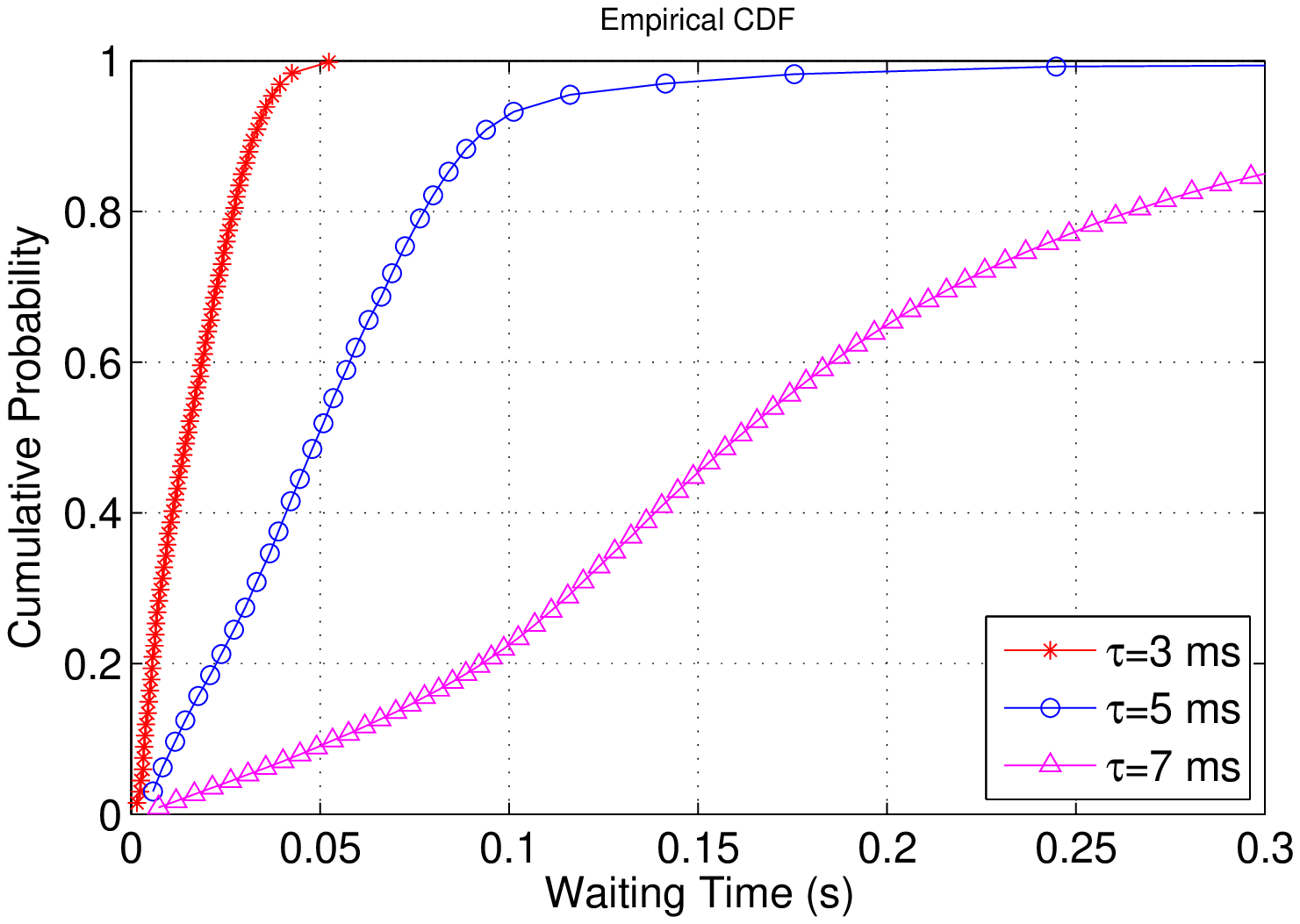}%
		\caption{}%
		\label{subfig2c}%
	\end{subfigure}%
	\caption{Waiting time CDF as a function of (a) $\alpha$, (b) $Pr_{check}$, and (c) $\tau$.}
	\label{fig2abc}
\end{figure*}

We use OMNeT++~\cite{omnet} and the IEEE 802.11p module from Veins~\cite{veins} to simulate our scheme and analyze the system performance. Let \emph{waiting time} be the total time a received message waits in the queue before its verification. Our scheme achieves low delays, which would not have been possible for the standard approach (\acl{FCFS} with verification of all messages, referred next as \emph{baseline}).
\begin{table}[htp]
	\caption{System Parameters (\textbf{\emph{Bold}} for Default Setting)}
	\centering
	\renewcommand{\arraystretch}{1.10}
	\begin{tabular}{ l | c }
		\hline \hline
		$N$ & 15, 20, 25, \textbf{\emph{30}}, 35, 40 \\\hline
		$Pr_{check}$ & \emph{\textbf{0.2}}, 0.4, 0.6 \\\hline
		$\alpha$ & 3, 4, \textbf{\emph{5}}, 6 \\\hline
		$\tau$ & 3, \textbf{\emph{5}}, 7 $ms$ \\\hline
		$\gamma$ & \textbf{\emph{10}} $Hz$ \\\hline
		\hline
	\end{tabular}
	\renewcommand{\arraystretch}{1}
	\label{table:parameter}
\end{table}

\textbf{Simulation Settings:} Table~\ref{table:parameter} shows the system parameters and values used in the simulation. We consider a $200\ m\times200\ m$ square area. The node we evaluate is placed at the center of the area and its neighbors (the rest of the nodes) are uniformly placed in the area. We consider a bit rate of 6 $Mbps$, with 300 $bytes$ length for each \ac{CAM} (including a signature and a certificate). In addition, we increase the payload length based on the number of hashes (80 $bit$ for each) included, thus communication overhead of $80\cdot\alpha$ $bits$. This amounts to extra communication delay in the order of 0.1 $ms$ per \ac{CAM}, considering the values we use for $\alpha$. Certificate omission~\cite{calandriello2011performance} can be used to decrease message verification delay. However, we do not explicitly address it in our simulation; rather, we assume the message verification delay, on average, has a deterministic value $\tau$ (including all operations, such as hash computation and queue search). For each simulation setting, we perform 5 randomly seeded experiments of 2 $min$ and average over these 5 runs. The bold values in Table.~\ref{table:parameter} are the default ones used in our simulation. For example, $N$ is the parameter we examine in Fig.\ref{subfig1a}, with the rest of the parameters having the default values: $Pr_{check}=0.2$, $\alpha={5}$, $\tau=5$ $ms$ and $\gamma=10$ $Hz$. The default values of $\tau$  and $\gamma$ are typical values based on the literature (e.g.,~\cite{calandriello2011performance,cam}).

\textbf{Simulation Results:} Fig.~\ref{subfig1a} shows the waiting time CDF as a function of $N$. We also evaluate the baseline scheme for $N=15$: maximum sustainable neighbor size (with such queue) is around 20, as only 200 $msg/s$ can be verified with $\tau=5$ $ms$. Fig.~\ref{subfig1a} shows the baseline scheme ($N=15$) performs similarly to our scheme with $N=20$. In addition, around 90\% of the messages have waiting times less than $0.3$ $s$ with $N=40, 45$, which is a significant improvement considering the queue would not even be stable for the baseline scheme.

Fig.~\ref{subfig1b} illustrates waiting times for the default setting over the simulation time. We see spikes, but overall, waiting times are stable and do not increase as simulated time progresses. In Fig.~\ref{subfig1c}, we see that as the number of neighbors increases (thus, higher message reception rate), more messages need to be verified based on the peer-provided verification results. This implies a higher risk of accepting bogus messages in a malicious environment. However, on the other hand, the probability of revealing (thus revoking) malicious nodes also increases as the neighbor density increases (Sec.~\ref{sec:analysis}).

We evaluate the waiting time for different $\alpha$ values (Fig.~\ref{subfig2a}). We discover a threshold for $\alpha$, above which peer-provided verification results cannot help: with $\alpha=7$, the waiting time CDF almost overlaps with that for $\alpha=6$. However, we can expect a higher threshold for $\alpha$ with larger $N$. $Pr_{check}$ has an impact on the probability of bogus message (thus, malicious node) detection and the waiting time distribution: more signatures need to be verified for higher $Pr_{check}$. As shown in Fig.~\ref{subfig2b}, almost 80\% of the messages have waiting times less than $0.8$ $s$ even with $Pr_{check}=0.6$. Under this setting, there is even a high probability that malicious nodes can be detected locally/independently, since more than half of the messages are checked. Fig.~\ref{subfig2c} shows the impact of message verification delay on waiting time. In our simulation, we found with $\tau=8$ $ms$, the queue is not stable anymore and the queue size goes to infinity. However, it is already a significant improvement considering only around 15 neighbors can be sustained for the baseline scheme with $\tau=7$ $ms$.

We do not consider deadlines for \acp{CAM} here, but we can infer those; e.g., for the default setting, with a 0.1 $s$ deadline, we can expect more than $90\%$ of the received \acp{CAM} to be verified (Fig.~\ref{subfig1a}). Again, this would not have been possible for the baseline scheme, which can sustain around 20 neighbors in the same setting.

\section{Conclusions}
\label{sec:conclusions}

We demonstrated how our cooperative message verification scheme could enable secure \ac{VC} at network densities even double compared to those prior approaches could be workable for. Though this is achieved by trading off a tiny vulnerability window, we showed this can be harmless. In addition, our scheme is orthogonal to all prior optimizations and could complement them.

\bibliographystyle{abbrv}
\bibliography{references.bib}
	
\end{document}